\newif\ifAlpha \Alphafalse

\newif \ifhyperlinks    \hyperlinkstrue

\newif \ifDraft         \Drafttrue
\newif \ifAnon	\Anontrue
\newif\ifFinal 		\Finalfalse
\newif\ifPreprint	\Preprintfalse

\Draftfalse
\Preprinttrue

\ifPreprint
  \Anonfalse
\fi

\ifFinal
  \Anonfalse
\fi

\documentclass[pageno,camera]{jpaper}
\newcommand{\asplossubmissionnumber}{454}

\ifPreprint
  \pretitle{\begin{center}\normalfont\Large\bfseries}
  \posttitle{\par\end{center}}
\fi

\usepackage{paralist}
\usepackage{times}

\ifAlpha
  \let \citep = \cite
  \let \citet = \cite 
  \usepackage{cite}
\else
   \usepackage[square,sort&compress,authoryear]{natbib}
  \renewcommand{\cite}[1]{\errmessage{Don't use cite, you probably want citep}}
\fi

\usepackage[normalem]{ulem}
\usepackage{subfig}
\newcommand{\autorefsub}[2]{\autoref{#1}\subref{#2}}

\usepackage{verbatim}           

\usepackage{xspace}             

\usepackage{balance}            

\usepackage{graphicx}
\setkeys{Gin}{keepaspectratio=true,clip=true,draft=false,width=\linewidth}
\graphicspath{{./imgs/}}
\newcommand{\code}[1]{\texttt{#1}}
\usepackage{tikz}
\usetikzlibrary{arrows}

\usepackage{hyperref}
\ifhyperlinks\else   
  \hypersetup{nolinks=true}
\fi

\begin{document}
  \sloppy

  \renewcommand{\sectionautorefname}{Section}
  \renewcommand{\subsectionautorefname}{Section}
  \renewcommand{\subsubsectionautorefname}{Section}
  \renewcommand{\appendixautorefname}{Appendix}
  \renewcommand{\Hfootnoteautorefname}{Footnote}
  \newcommand{\Htextbf}[1]{\textbf{\hyperpage{#1}}}

\title{An Evaluation of Coarse-Grained Locking for Multicore Microkernels}
\ifAnon 
  \author{\ifDraft Draft of \today, \pageref{p:lastpage} pages of 11 allowed.\fi}
\else
  \author{Kevin Elphinstone, Amirreza Zarrabi, Adrian Danis, Yanyan Shen, Gernot Heiser\\
    UNSW and Data61, CSIRO, Australia\\
    \href{mailto:kevin.elphinstone@data61.csiro.au}{\emph{firstname.lastname}@data61.csiro.au}}
\fi
\date{}

  \maketitle
  \urlstyle{sf}
  \thispagestyle{empty}
  \begin{abstract}
      The trade-off between coarse- and fine-grained locking is a well
    understood issue in operating systems. Coarse-grained locking
    provides lower overhead under low contention, fine-grained locking
    provides higher scalability under contention, though at the
    expense of implementation complexity and reduced best-case
    performance. 

    We revisit this trade-off in the
    context of microkernels and tightly-coupled cores with shared
    caches and low inter-core migration latencies. We evaluate
    performance on two architectures: x86 and ARM MPCore, in the
    former case also utilising transactional memory (Intel TSX). Our
    thesis is that on such hardware, a well-designed microkernel, with
    short system calls, can take advantage of coarse-grained locking
    on modern hardware, avoid the run-time and complexity cost of
    multiple locks, enable formal verification, and still achieve scalability comparable to
    fine-grained locking.
  \end{abstract}
\ifFinal
  \pagestyle{empty}
\fi

\section{Introduction}\label{s:intro}

Waste of processing power resulting from lock contention has been an
issue since the advent of multiprocessor computers, and has become a
mainstream computing challenge since multicores became
commonplace. Much research is directed to understanding and achieving
scalability to large numbers of processor cores, where lock contention
is inevitable and must be minimised \citep{Clements_KZTK_13}. \emph{It is
now taken as given that locks must be fine-grained}, ideally protecting
individual accesses to shared data structures, and that shared
data structures must be minimised, or, in the extreme case of a
multikernel \citep{Baumann_BDHIPRSS_09}, avoided altogether.

We observe that a discussion of scalability cannot be done without
taking into account operating system (OS) structure as well as
platform architecture.
Prior scalability work is typically performed in the context of a monolithic
OS that needs to scale to hundreds or thousands of
concurrent hardware execution contexts, and communication between
contexts measuring in the thousands of cycles. But 
monolithic systems are no longer all that matters, microkernels are finding
renewed interest due to their ability to reduce a system's trusted
computing base and thus its attack surface \citep{McCune_PPRI_08,
  Steinberg_Kauer_10, Heiser_Leslie_10, Zhang_CCZ_11, Klein_AEMSKH_14}.

In a monolithic system, such as
Linux, typical system call latencies are long, even compared to 
inter-core communication latencies in the 1000s of cycles. In contrast,
a well-designed microkernel is essentially a context-switching engine, with typical
syscall latencies in the hundreds of cycles
\citep{Heiser_Elphinstone_16}. In such a system, the cost of
cross-core synchronisation may be an order of magnitude higher than
the basic syscall cost. It therefore makes no sense to run a single
kernel image, with shared data structures, across such a manycore
machine. An appropriate design should share no data between cores
where communication is expensive, resulting in a multikernel design
\citep{Baumann_BDHIPRSS_09}.

However, the multikernel approach is not the complete answer
either. It presents itself to user-level as a distributed system,
where userland must explicitly communicate between nodes. This is not
the right model where communication latencies are small, eg.\ across
hardware contexts of a single core, or between cores that share an L2
cache, where they are of the order of tens of cycles, well below the
latency of a syscall even in a microkernel. In this context, explicit 
communication between nodes is more expensive
than relying on shared memory, and there is no justification for
forcing a distributed-system model on userland.

We therefore argue that, for a microkernel, the right model is one
that reflects the structure of the underlying hardware: a shared
kernel within a closely-coupled cluster of execution contexts, but
shared nothing between such clusters. The resulting model is that of a
\emph{clustered multikernel} \citep{vonTessin_12}.

A node in such a cluster puts scalability into a different context:
rather than to hundreds or thousands of cores, it only needs to scale
to the size of a closely-coupled cluster, no more than a few dozens of
execution contexts. Such a cluster matches another important category
of platforms: the now ubiquitous (and inexpensive) low-end multicore
processors deployed by the billions in mobile devices.

A microkernel for a closely-coupled cluster represents an area in the
design space markedly different from that of manycores, and it is far
from obvious that the same solutions apply. In particular, it is far
from obvious that fine-grained locking makes sense. In fact, the
typical systemcall latencies are not much longer than critical section
in other OSes. In that sense, for a microkernel, even a big lock 
is not much coarser than a fine-grained lock in a system like Linux --
as has been observed before, for a microkernel, a big lock may be fine-grained enough \citep{Peters_DEH_15}.

This discussion may seem academic at first, given that fine-grained
locking techniques are well-known and widely implemented, so why not
use them anyway? There are, in fact, strong reasons to stick with
coarse-grained locking as long as possible: Each lock acquisition has
a cost, which is pure overhead in the absence of contention. While
insignificant compared to the overall system-call cost in a system
like Linux, in a microkernel this overhead is significant.

More importantly, the concurrency introduced by fine-grained locking greatly
increases the conceptual complexity of code, and thus increases the
likelihood of subtle bugs that are hard to find
\citep{Lehey_01}, as painfully confirmed by our experience
implementing fine-grained locking in seL4. Furthermore, this
complexity is presently a show-stopper for formal verification,
which otherwise is feasible for a microkernel
\citep{Klein_AEMSKH_14}. 

Additionally, as Intel TSX \emph{restricted transaction memory} (RTM)
extensions become widely available, there is an opportunity to have
the complexity of coarse-grained locking and the performance of
fine-grained locking by using RTM to elide coarse-grained locks
\citep{Rajwar_Goodman_01}.

We therefore argue that it is important to understand the performance
impact of a big-lock design, which maximises best-case performance,
minimises complexity and eases assurance.
To this end we conduct a detailed examination of the scalability of the seL4
microkernel on closely-coupled clusters on two vastly
different hardware platforms (an x86-based server and an ARM-based
embedded processor) under different locking regimes. We make the
following contributions:
\begin{compactitem}
\item We estimate the theoretical performance of a coarsely
  synchronised (big-lock) microkernel using queueing theory (\autoref{s:model}).
\item We validate the queueing model experimentally, and at the same
  time identify modifications to the microkernel to achieve near
  theoretically optimum performance (\autoref{s:ubm}).
\item We compare big-lock and fine-grained-lock
  implementations of the seL4 microkernel and evaluate those on
  closely-coupled cores on two architectures (ARM and x86), in
  contrast to the usual approach of aiming for high-end scalability
  across loosely-coupled cores (\autoref{s:eval}).
\item We present (on x86) the first use of hardware transactional
  memory that places the majority of the kernel into a single
  transaction for concurrency control, and we compare it with
  locking (\autoref{s:eval}). 
\end{compactitem}

We show that the choice of concurrency control in the kernel is
clearly distinguishable for extreme synthetic benchmarks \autoref{s:ubm}). For a
realistic, system-call intensive workload, performance differences are
visible, and coarse grain locking is preferred (or the equivalent
elided with hardware transactional memory) over extra kernel
complexity of fine-grained locking (\autoref{s:redis}).

\section{Background}\label{s:back}

\subsection{Locking granularity}

The best locking granularity is determined by a trade-off involving
multiple factors. As long as there is no contention, taking and
releasing locks is pure overhead, which is minimised by having just a
single lock, the \emph{big kernel lock} (BKL). Each lock adds some
overhead which degrades the best-case (i.e.\ uncontended) performance.

As long as the total number of locks is small, this baseline overhead
is usually small compared to the basic system-call cost. However, on a
well-designed microkernel, where system calls tend to be very short
(100s of cycles) this overhead might matter.

Fine-grained locking can significantly
reduce contention, if it enables unlocked execution of the majority of
code. In a BKL kernel, contention can be expected to be noticeable as
soon as the \emph{hold time} (fraction of time spent inside the
kernel, also referred to as \emph{kernel time}) is
not small compared to the \emph{pause time} (fraction of time spent in
user mode).

The amount of kernel time depends on the profile of system calls
executed, and thus on the workload. On a monolithic kernel, most
system services are provided by the kernel, especially I/O, and
consequently I/O-intensive workloads tend to have high kernel time. On
a microkernel, system services are provided by server processes
running in user mode, and the kernel provides communication between
clients and servers. On a well-designed microkernel, such as the ones
of the L4 family, kernel time is dominated by context
switches \citep{Liedtke_95}. The total number of kernel calls is higher
than in a monolithic kernel (at least twice as high, as every server
invocation invokes the kernel twice) but the average system-call
latency is a tiny fraction of that of a monolithic kernel.

Hence, a BKL is a more credible design for a microkernel than for a
monolithic OS, at least for a closely-coupled cluster of execution
contexts, where intra-cluster communication latencies are low (eg.\
due to a shared cache). As explained in the introduction, this does
not prevent the kernel from use in manycores, but on such hardware,
kernel state should not be shared across clusters, resulting in the
clustered multikernel design. It is the design of a kernel for such a
cluster which we explore in this paper. To avoid drawing invalid
conclusions from idiosyncrasies of a particular platform, we examine two very
different architectures: an x86-based server processor and an
ARM-based processor aimed at embedded devices, especially
smartphones. 

\subsection{x86 platform}
\label{s:x86plat}

\begin{table}[htb]
  \centering
  \begin{tabular}{|l|l|r|r|r|}\hline
    &
    \multicolumn{1}{c|}{Memory} & 
    \multicolumn{1}{c|}{Local} & 
    \multicolumn{1}{c|}{Intra-} & 
    \multicolumn{1}{c|}{Inter-}  \\

    \multicolumn{1}{|c|}{Platform} &  
    \multicolumn{1}{c|}{Level} &  
    \multicolumn{1}{c|}{core} & 
    \multicolumn{1}{c|}{socket} & 
    \multicolumn{1}{c|}{socket}  \\\hline
    x86 & L1     &   4 &   115 & 218 \\
        & L2     &  12 &   105 & 208 \\
        & L3     &  44 &    44 & 163 \\
        & Memory & 185 &   185 & 265 \\\hline
    ARM & L1     &   4 &   17 & N/A \\
        & L2     &  26 &   28 & N/A  \\
        & Memory & 140 &   N/A & N/A \\\hline
  \end{tabular}
  \caption{Memory and cache access latency in cycles.}
  \label{tab:latency}
\end{table}

As an x86 platform we use a server-class Dell Poweredge R630 fitted
with two Intel Xeon E5-2683 v3 processors. These are a 14-core
processors with a base clock rate of 2.0\,GHz and two hardware threads
each, giving 28 hardware threads per processor. Thus the machine has
total of 56 hardware threads across the two CPU sockets. While not
officially supported, the microarchitecture features Intel's TSX
implementation of \emph{restricted transactional memory} (RTM), which
we describe further in \autoref{s:rtm-arch}.

The processor features three levels of cache. Each core has private L1
instruction and data caches, each 32\,KiB in size and 8-way
associative. Each core furthermore has a private, non-inclusive, 8-way
256\,KiB L2 cache. The last-level cache is 35MiB, consisting of a
2.5MiB slice per core. 

\autoref{tab:latency} shows our measured memory latency, cache
access latency, and latency of data transfer between cores on the
same socket, and across separate sockets. The measurements were obtained using
code derived from \emph{BenchIT}, and the results are reasonably
consistent with measurements obtained by the
authors \citep{Molka_HSN_15}, noting the differing clock rates of the
system under test. One should also note that these results are 
sensitive to the distance between cores and thus will vary depending
on the specific cores involved.

\subsection{ARM platform}
\label{s:armplat}

Our ARM platform is the Sabre Lite, which is based on a
Freescale \mbox{i.MX 6Q} SoC, featuring a quad-core ARM Cortex-A9
MPCore processor \citep{imx6}.

The cores run at a 1\,GHz clock rate and have private, split L1
caches, each 4-way-associative and 32\,KiB in
size. The cores share a 1\,MiB, unified, 16-way-associative L2
cache, which is the last-level cache. We ported the microbenchmarks
from \emph{BenchIT} to the ARM platform and obtain the results in
\autoref{tab:latency}.

Compared to x86, the ARM has much lower latency of data transfer
between caches, and the latency is unaffected by distance between the
cores.

\subsection{Intel TSX\label{s:rtm-arch}}

TSX provides 4 new instructions: \code{XBEGIN}, \code{XEND},
\code{XTEST} and
\code{XABORT}. Code successfully executed between \code{XBEGIN} and \code{XEND}
instructions will appear to have completed atomically, and is thus
called a transactional region. If there are any memory conflicts
during the execution of the transactional region, the transaction
will abort and jump to the instruction specified by the \code{XBEGIN}.
A program can explicitly abort a transaction by issuing an
\code{XABORT} instruction. \code{XTEST} returns whether currently
executing within a transactional region.

TSX takes advantage of existing cache coherency protocols, to identify
sets of cache lines written to and read by different cores on the
CPU. This has two important consequences: memory conflicts are
captured at a cache-line granularity, and transactions are constrained
by the size of the L1 and L2 caches. The mutated state must fit inside
L1 cache, and the accessed state must fit inside the L2 cache
\citep{Hasenplaugh_NS_15}.  The consequence is that it is probably not
feasible to wrap a complete monolithic kernel into an RTM transaction,
as it is unlikely to fit within the L1 and L2 caches.

Owing to the implementation of TSX, the RTM lock logically protects
a dynamic set of individual L1 and L2 cache lines, and as such is a fairly extreme case of
fine-grained locking, which should result in much reduced contention
(assuming a sane layout of kernel data structures). 

Note that an RTM transaction is not guaranteed to complete, even when the 
transaction is small enough and has no memory conflicts. A variety of 
(hardware-implementation specific and frequently unspecified) scenarios can result in an abort. Of
particular interest to our work are certain interactions on specific 
registers that trigger aborts, but are clearly unavoidable when executing OS code.

Given transactions have no guarantees of progress, the developer must
ensure that there exists a fallback method of synchronisation that ensures
progress in the presence of repeated aborts. We use the commonly implemented
technique of falling back on a regular lock for the code fragment in
the case of repeated aborts. To avoid races between transactions and locks,
our transactions test
the lock upon entry to an RTM section, to ensure the lock is free
and force it into the read set of the transaction. A change in lock state
by a competing thread will trigger the desired abort, and allow the
section to synchronise via the lock.

\section{Microkernel Implementation}\label{s:impl}

\subsection{seL4}
As we use seL4 as our microkernel testbed, we will now summarise its
relevant features, \citet{Klein_AEMSKH_14} presents more details.
seL4 is event-based, with a single kernel stack. To aid verification, seL4 uses a two-phase system call structure, where the first phase confirms the
pre-conditions required for system call execution, and the second
phase executes the system call without failure. Blocking operations are handled by
re-starting the system call and thus re-confirming the preconditions
prior to continuing execution.

The kernel executes with interrupts disabled. This concurrency-free
design has traditionally been used in L4 kernels in order to achieve
high best-case performance, and has been used on other systems as well
\citep{Ford_HLMT_99}. With formal verification it becomes as
necessity, as it is for now infeasible to verify concurrent kernel
code.

The kernel features some long-running operations resulting from the
destruction of kernel objects that may have derived objects. In order
to achieve usable interrupt latencies, it has explicit
preemption points, where the kernel polls for pending interrupts, and
restarts the operation if there are any \citep{Blackham_SH_12}. The
restart allows interrupts to be
triggered from outside the kernel, prior to continuing the original operation. 

seL4 supports the traditional L4-style synchronous (rendez-vous)
message passing IPC with a payload of up to a few hundred bytes. IPC operates via port-like objects called \emph{endpoints}.  In
addition, the kernel provides \emph{notifications} with semantics similar
to binary semaphores. 

\subsection{Big kernel lock}
\label{s:bkl}

The BKL is the natural, minimal extension of the existing seL4 design
to multicores, as it is easy to implement and mostly preserves the
in-kernel assumption of no concurrency. The kernel entry and exit
code, which saves and restores the user-state to a per-core kernel
stack and sets up safe kernel
execution, remains outside of the BKL, while the rest of
the kernel is protected by the BKL. 

This design is not entirely sufficient -- the following invariant,
used in the verification, no longer holds on a multicore kernel, even
when the BKL is held:
\begin{quote}
  Except for the currently executing thread's TCB and page table, all
  other TCBs and page tables are quiescent, and can be mutated or
  deleted.
\end{quote}

User-level code executing on other cores implicitly depends on the
running thread's TCB and page table to transition to kernel-mode via
the kernel entry code to compete for the BKL. The invariant therefore
no longer holds. We address this by modifying the kernel to ensure
remote cores are not dependent on any TCB or page-table undergoing
deletion.  

We modify our prototype to keep a bitmap of cores that have seen a
specific page table in the page table itself, and IPI only those cores
to trigger the remote core to enter the kernel idle loop (which has a
permanently allocated TCB and page-table), and also to shoot down the TLB. A TCB
can be identified as active via the CPU affinity in the TCB itself
combined with the per-core \emph{current thread} pointer of the remote core, in
which case the TCB is handled in a similar manner to the page table.

This design, which is partially driven by the existing event-driven
code base, is a valid design choice thanks to the short duration of
most system calls in the microkernel; it would result in poor
scalability on any other kind of system.

The only other required change is introducing per-core idle
threads. However, in order to minimise inter-core cache-line
migrations, we also introduce per-core scheduler queues in addition to
the current-thread pointers, even though access is serialised by the
BKL. This partitioned scheduling implies that threads can only migrate
between cores if explicitly requested by the user, which is consistent
with seL4's general philosophy of having all resource management under
user control (and also helps reasoning about real-time properties).

To reduce contention (and enable the use of transactional memory, see
\autoref{s:rtm-arch}) we further minimise the amount of locked code by moving
context-switch-related hardware operations after the BKL release,
which has the benefit of reducing the critical section length.

We use a CLH lock, as scalable queue lock \citep{Craig_93}, to
synchronise the BKL kernel variant.

\subsection{Fine-grained locking}
\label{s:fine-grained}

To compare the coarse-grained BKL with more complex but more scalable
fine-grained locking, we first replace the BKL with a big reader lock
\citep{corbet:url}. The lock allows all reader cores to proceed in
parallel as they access only local state to obtain a read lock. 

In our present prototype we use a single write lock around the the
non-IPC-related kernel code paths. These code paths, generally dealing
with resource management, are infrequently executed, compared to IPC
and interrupt handling, and as such not performance critical. This
allows us to avoid significant code changes without affecting overall
performance.

This design allows us to gradually migrate the kernel code out of
the writer lock into the reader lock.  As long as deallocation of
kernel objects remains inside the writer lock, memory safety is
retained while holding the reader lock. Moving code into the reader
lock exposes the \emph{contents} of the objects to concurrency for
improved scalability, which can then be protected using individual
fine-grained locks.

IPC mutates the state of TCBs, endpoints, and (potentially) the
scheduler queues (depending on whether optimisations apply that avoid
queue updates during IPC \citep{Heiser_Elphinstone_16}). We add ticket
locks to each of these data structures for synchronising IPC within
the reader lock. A typical IPC now involves the kernel
reader lock, two TCB locks, and one endpoint lock. Lock contention
during IPC is limited to cases where IPC involves a shared destination
or endpoint, or general contention with the kernel writer
lock. Independent activities performing IPC on independent cores
result in no lock contention. We avoid deadlocks
by identifying the affected TCBs prior to locking (made
possible by memory safety provided by the reader lock), and then
locking them in order of their memory addresses.

\subsection{Hardware transactional memory}
\label{s:rtm}

The TSX extensions, combined with the small size of the kernel, allow
us to optimistically execute the majority of the code without
concurrency control. This is analogous to taking the BKL kernel
variant and speculatively eliding the lock
\citep{Rajwar_Goodman_01}. The event-based design of the kernel is an
important enabler for lock elision as it avoids blocking.  We bracket
almost the entire kernel with the transaction primitives shown in
\autoref{f:code}. The somewhat simplified code is self explanatory,
except the `L' argument to \code{\_xabort()}, which is returned as the
status at \code{\_xbegin()} to distinguish between abort types.

\begin{figure}[htb]\small
\begin{verbatim}
beginTransaction() {
  while ((status = _xbegin()) != 
         _XBEGIN_STARTED ) {
    txnAttempts++;
    if (txnAttempts >= 
        RTM_ATTEMPTS_THRESHOLD) {
      break; /* Give up */
    }
    /* wait for lock freed before retrying txn */
    while(LockTest()); 
  }
  if (status == _XBEGIN_STARTED) {
    if (LockTest()) {  /* not free */
      _xabort('L');
    }
  } else {
    lockAcquire(); /* BKL fall-back */
  }
}

endTransaction() {
    if ( (txnInside = _xtest()) ) {
        _xend();
    } else {
        lockRelease();
    }
}
\end{verbatim}
  \caption{Kernel transaction pseudo code.}
  \label{f:code}
\end{figure}

In addition to the changes described in \autoref{s:bkl}, we need to
move any TSX-specific abort-triggering CPU operations after the transaction. 
Many of those do not occur in seL4, as most aborting
operations are typical for device drivers, which are user-level
programs in seL4. The remaining problematic operations are:
\begin{compactitem}
\item context-switch-triggered page-table register (CR3) loading and segment-register loading;
\item IPI triggering for inter-core notifications;
\item interrupt management for user-level device drivers, which consists of masking and acknowledging interrupts prior to return to the user-level handler.
\end{compactitem}

The key insight here is that it is safe to move these operation
outside of the transaction, because the two-phase kernel ensures the
system call which requires these operations is guaranteed to succeed
once the execution phase is entered, and that these operations are
local to a core and thus are not exposed to concurrent access from
other cores. Note that preemptions during this code section are
prevented since the kernel runs with interrupts disabled.

\section{BKL multicore scalability}\label{s:model}

In this section we use queuing theory to model the scalability of a
BKL microkernel. The theoretical model provides us with a method to
estimate best-case performance for a workload parametrised by the rate
the lock can be serviced ($\mu$), the arrival rate of the lock
requests ($\lambda$, i.e. system call rate), and the number ($n$) of
cores in the machine. An estimate of best-case performance provides a
theoretical reference point to target.

\subsection{Modelling contention}

We employ the \emph{machine repairman} queuing theory model. The model
is historically based on machine failures in a factory (characterised
by a failure rate) combined with waiting for a single repairman (the
service rate). In our case the model corresponds to the
arrival rate of lock requests combined with the service rate of the
lock itself.

\begin{figure}[htb]
  \centering
\resizebox{\hsize}{!} {
  \begin{tikzpicture}[->,>=stealth',shorten <=
    1pt,auto,node distance=1.9cm,
    thick,state/.style={font=\footnotesize,minimum size=25pt,circle,draw},dots/.style={minimum size=25pt}]
    \node[state] (1) {$0$};
    \node[state] (2) [right of=1] {$1$};
    \node[dots] (d) [right of=2] {$\cdots$};
    \node[state] (3) [right of=d] {$n-1$};  
    \node[state] (4) [right of=3] {$n$};
    \path[every node/.style={font=\sffamily\small}]
    (1) edge [bend left] node  {$\lambda_0$} (2)
    (2) edge [bend left] node {$\mu_0$} (1)
    (2) edge [bend left] node  {$\lambda_1$} (d)
    (d) edge [bend left] node {$\mu_1$} (2)
    (d) edge [bend left] node  {$\lambda_{n-2}$} (3)
    (3) edge [bend left] node {$\mu_{n-2}$} (d)
    (3) edge [bend left] node  {$\lambda_{n-1}$} (4)
    (4) edge [bend left] node {$\mu_{n-1}$} (3);
  \end{tikzpicture}
  }
  \caption{The machine repairman model.}
  \label{fig:mrm}
\end{figure}
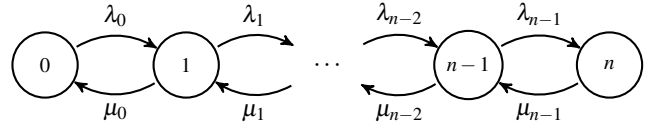

The model for an $n$-core multiprocessor has $n+1$ states, representing
the number of cores holding or waiting to acquire the kernel lock, as
illustrated in \autoref{fig:mrm}.

The model assumes the rate, \(\mu\), of servicing the lock, to be
independent of the number of cores queued, i.e.\
$\mu_{k} = \frac{1}{s}$, where $s$ is the average service time of the
lock. It further assumes that the rate of arrivals, \(\lambda\), is
proportional to the number of cores not already waiting for the lock,
i.e.\ $\lambda_{k} = \frac{n-k}{a}$, where $a$ is the average
inter-arrival time for a single core in the absence of contention.

In a steady state, the rates of lock acquisitions and lock releases
must be balanced. 
If  $P_i$ is the probability  of being in state \(i\), this means
$P_{k}\cdot \lambda_{k} = P_{k+1}\cdot \mu_{k}$. From this we can
derive the probabilities as

\begin{equation}
  \label{eq:pk}
  P_{k}=P_{0}\cdot \frac{s^{k}n!}{a^{k}(n-k)!}\ .
\end{equation}

The system must always be in one of these states, \(\sum_{i=0}^{n}P_i
= 1\), from which we can obtain

\begin{equation}
  \label{eq:p0}
  P_{0}= \frac{1}{\sum_{i=0}^{n}(\frac{s^{i}n!}{a^{i}(n-i)!})}\ ,
\end{equation}

and ultimately

\begin{equation}
  \label{eq:sspk}
  P_{k} = \frac{ \frac{s^{k}}{a^{k}(n-k)!}}{
    \sum_{i=0}^{n}(\frac{s^{i}}{a^{i}(n-i)!})}\ .
\end{equation}

From this we can compute the expected queue length as
\begin{equation}
  \label{eq:qlen}
 w = \sum_{i=0}^{n}i P_{i}\ ,
\end{equation}

and lock throughput is
\begin{equation}
  \label{eq:tput}
  \mu(1-P_{0}) = \frac{1-P_{0}}{s}\ .
\end{equation}

\subsection{Model Assumptions and Kernel Design}

The queueing model assumes that the average rate of serving the lock
is independent of the queue depth. This is not true for non-scalable
locks or in the case of mutating shared kernel state
\citep{Boyd-Wickizer_KMZ_12}. We satisfy these assumptions by avoiding
shared mutable state for unrelated kernel system calls through per-core
data structures, and using the scalable CLH  lock.

In addition, peak throughput is inversely proportional to lock service
time. Hence, moving as much code out of the lock as possible, in
particular the expensive local
hardware operations (such as triggering of IPIs, and page table
register updates), will improve scalability.

\section{Evaluation}\label{s:eval}

To evaluate our multicore microkernel variants, we use two IPC
microbenchmarks and a server-style macrobenchmark, as described in the
following sections. The platforms under test have been already been
described in Sections~\ref{s:x86plat} and \ref{s:armplat}.

\subsection{Microbenchmarks}\label{s:ubm}

\subsubsection{Single-Core IPC microbenchmarks}\label{s:1c-pingpong}

IPC performance is a key contributor to overall system performance
in microkernel-based systems, and optimising IPC performance has a long
history in the L4 community \citep{Heiser_Elphinstone_16}. The
traditional benchmark for best-case IPC performance is ``ping-pong'':
a pair of threads on a single core does nothing other than sending
messages to each other. This allows us to asses the basic cost of our
lock implementations, i.e.\ the pure acquisition and release cost,
without any contention.

\begin{figure}[htb]
  \centering
  \subfloat[x86]
  {\includegraphics[width=0.58\columnwidth]{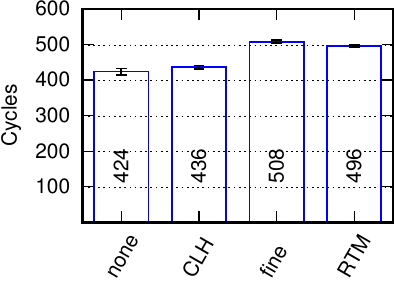}\label{f:x86-ipc}}%
  ~~%
  \subfloat[ARM]
  {\includegraphics[width=0.34\columnwidth]{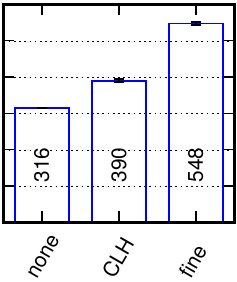}\label{f:arm-ipc}}
  \caption{\label{f:ipc}Raw one-way IPC cycle cost for different
    seL4 locking mechanisms. Error bars indicate standard
    deviations.}
\end{figure}

The figure shows that on x86, the overhead of a single CLH lock (BKL) is
approximately 3\% compared to a baseline
uniprocessor kernel with no concurrency primitives (``none''). With
fine-grained locking, however, the overhead is 20\%.  The overhead of
uncontended transactions is 17\%.

On the ARM, the cost of a single lock is significantly higher with 23\%,
while for fine-grained locking it is over 70\%.
The higher
synchronisation costs on the ARM processor relate to its partial-store-order memory
model. It requires memory barriers (\code{dmb} instructions) to
preserve memory-access ordering. In our experience, the barriers cost
from 6 cycles up to 19 cycles depending on micro-architectural
state. Our implementation of CLH executes 6 barriers on this
benchmark, while 16 are needed with 
fine-grained locking. These barriers explain most of the overhead.

The significant cost of
fine-grained locking provides a motivation for sticking with the BKL
as long as possible, even if verification tractability was no issue.

\subsubsection{Multicore IPC BKL microbenchmarks}

\begin{figure*}[t]
  \subfloat[Scalability measurements and fits for 358 cycles service time.]
  {\includegraphics[width=0.525\hsize]{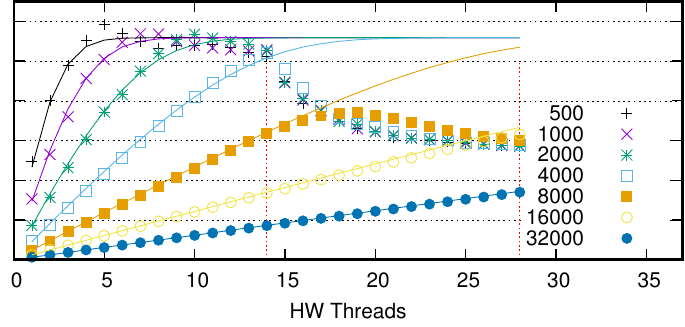}%
    \label{fig:queueing}}
  \hspace{\fill}
  \subfloat[Scalability bound prediction for service times of 323 cycles (upper
  lines) and 613 cycles (lower lines).]
  {\includegraphics[width=0.47\hsize]{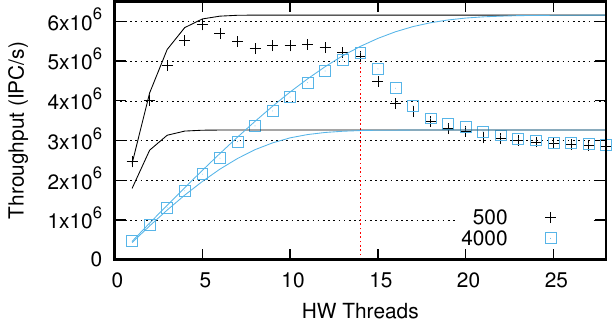}%
  \label{fig:bounds}}
  \captionof{figure}{Total IPC throughput for varying parallelism and delay
    times (cycles). Points are measurements, lines are queueing model fits.}
  \label{fig:model}
\end{figure*}

To explore scalability and experimentally validate the queueing model,
we extend the single-core ping-pong to multiple cores.
Specifically, we run a copy of ping-pong on each hardware
thread, with all hardware threads executing completely independently
and unsynchronised. We use the BKL kernel on the x86 platform.

We add
an exponentially distributed random delay between receiving
and replying to IPC, for each ping-pong pair. The delay varies
between an average of 500 and 32,000 cycles, in powers of 2, to create
seven individual microbenchmarks that simulate a work load from an
extreme system call intensive workload to a relatively compute-bound
workload.

The benchmark is \emph{embarrassingly parallel} ensuring that limits
of scalability are related to our kernel design and implementation,
and not the benchmark itself. The number of hardware threads (cores in
this case) varies between 1 and 28.

This benchmark produces extreme contention on the kernel (for low
delay values). However, \emph{none of the kernel data structures are
  contended}, as each hardware thread's pair of software threads
accesses disjoint kernel objects (TCBs and IPC endpoints) during their
syscalls.  Hence, while expecting contention on the BKL, fine-grained
locking and RTM can be expected to scale perfectly.

For each delay and core-count parameter pair, the benchmark consists
of a two second warm-up, followed by sampling total IPCs during a one
second interval to give total IPC throughput per second.

\autorefsub{fig:model}{fig:queueing} plots the resulting overall IPC
throughput for varying number of hardware threads.  Each point
represents one measurement for a particular delay time (identified by
symbol and colour). The vertical dotted lines shows where
the cores are split across the two CPU sockets. The results have
negligible variance.

For the runs with an average delay of 2000 cycles we perform a
least-squares regression of the queueing model using only the points
for the first 14 hardware threads (i.e.\ within one socket). The
regression yields a service time of 358 cycles, and an average
delay of 1999 cycles, with $R^2=0.99$, meaning that the intra-core
results are explained by the model if the service time is 358 cycles.

We use this service time to predict throughput for all other values of
the delay parameter, resulting in the solid lines in the graph. We can
see that in all cases, these fit the observed throughput values very
well for at least 14 threads (i.e.\ the model explains the
intra-socket behaviour well).

The model breaks
down once cores of the second socket are involved, except for the
highest delay times. This is
unsurprising, as with multiple sockets, the assumption of a fixed
service time no longer holds, as transfer times for the cache line
holding the lock now depend strongly on locality. We confirm this by
instrumenting the lock: The average holding time is 164 cycles for a
single core. For four or more cores, the observed holding times
vary between 323 and 613 cycles.\footnote{The
  average holding time of the lock is strongly influenced by the
  average transfer time of the cache line. Given transfer cost is zero
  for re-acquisition of the lock by the same core, low core counts
  have higher probability of re-acquisition, and thus unrealistically
  low average holding times for the general case.}

In \autorefsub{fig:model}{fig:bounds} we repeat the experimental results of
\autorefsub{fig:model}{fig:queueing} for two delay values, 500 and
4,000 cycles. We also show the model prediction for 323 and 613
cycles, the minimum and maximum holding times we observed for thread
counts of four or more. We can see that the results are upper-bounded
by the 323-cycle curve, and, as long as only one socket is
used, remain close to the bound. Once the
second socket is used, the lines quickly approach the lower line,
corresponding to the higher service time, which remains a lower bound.

\subsubsection{Observations about the model}

The queueing model accurately predicts experimental results where the average
lock holding time is stable. Where the lock holding time is variable,
it can be used to predict a performance range. The model enables
prediction of where the \emph{knee} of the performance curve occurs
for a given lock holding time and average delay between system calls,
assuming the absence of other application-related limiting factors.

The lock-holding time range for the microkernel varies from an average
164 cycles on a single core, to approximately 300 cycles within a
socket, to 600 cycles distributed across sockets. Thus the lock
holding time is dominated by the architectural cost of cache line
transfer for the lock. This is indirect confirmation that our
microkernel is indeed scalable in the sense of not sharing any mutable
state across cores except for the lock itself. It also implies that
any improvement in reducing lock holding time on a single core will
have only a modest effect on overall scalability due to the high
architectural costs on the Xeon.

The model and experiments show that a workload running on the
microkernel with an average delay between system calls on each core of
4000 cycles would scale to 14 cores, i.e. a single CPU socket. An
average delay time of 16000 cycles is needed for a workload to scale
across both sockets. The results support our hypothesis that a big
lock will scale as long cores are closely coupled.

These results are readily applicable in general to conservative locks
protecting \emph{potentially} contended data, that rarely contend in
practice. We also note for the following section that these results are
readily applicable to the abort path in the RTM variant of the
microkernel.

\begin{figure*}[t]
  \subfloat[x86, average delay 4,000 cycles.]
  {\includegraphics[width=0.62\hsize]{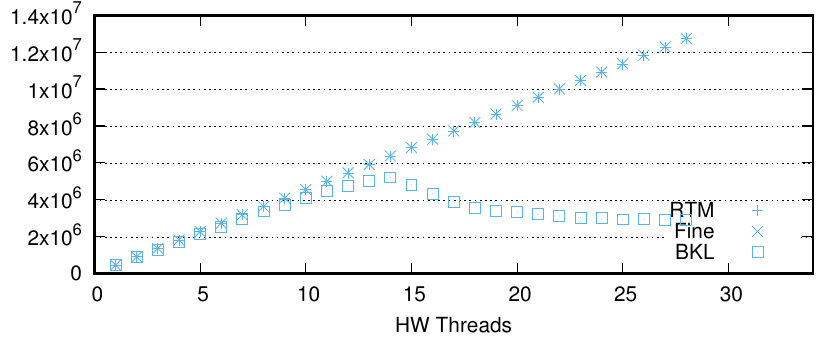}%
    \label{f:ipc_x86_tput}}
  \hspace{\fill}
  \subfloat[ARM, average delay 0 and 500 cycles. Lines are for clarity
  (no fit).]
  {\includegraphics[width=0.35\hsize]{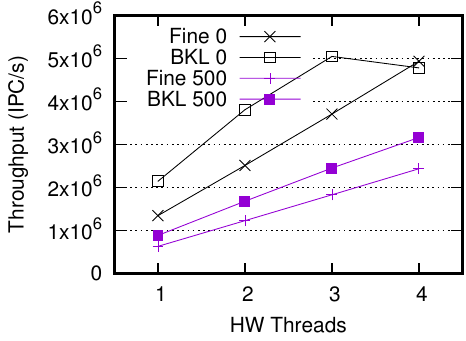}%
    \label{f:ipc_arm_tput}}
  \captionof{figure}{Total IPC throughput for varying parallelism and
    different locking implementations.}
  \label{f:ipc_variants}
\end{figure*}

\subsubsection{Locking variant evaluation}

To compare different lock variants, we run on x86 with a 4,000-cycle average
delay between system calls, as this is just above the scalability limit of
the BKL across a whole socket.

\autorefsub{f:ipc_variants}{f:ipc_x86_tput} shows that
RTM behaves identically to fine-grained locking. This is
expected: as explained in \autoref{s:rtm-arch}, RTM is logically an
extreme case of fine-grained locking, and the baseline lock overhead
is the same as for the fine-grained locks according to
\autoref{f:ipc}.

The BKL variant serialises the IPC path across all the available hardware
threads, thus hits the knee in the performance curve as predicted by
the queueing model. As also expected, it converges on a lower performance plateau as
the benchmark spans the sockets.

On ARM (\autorefsub{f:ipc_variants}{f:ipc_arm_tput}), where intra-core
cache line migration costs are very low (in terms of cycles), 
we chose an agressive 500-cycle average delay. In addition, we run the
pathological zero-delay case, where the system call rate is only
limited by the user-level stubs and cost of the system call itself. We
see the higher overhead for fine-grained locking is readily visible,
with the BKL variant outperforming the fine-grained variant for the
fairly extreme case of the 500 cycle average delay time, with perfect
scalability to all four cores. It takes the
unrealistic minimal delay for the BKL variant to plateau at 3 cores,
allowing the fine-grained variant to exceed BKL throughput. We can
expect the BKL to scale significantly beyond the size of our quad-core
machine for realistic workloads.

\subsection{Redis-based Macrobenchmark}\label{s:redis}

In order to assess BKL
scalability, and the significance of the overheads of the fine-grained
schemes, we look for a ``realistic worst-case'' scenario, i.e.\ a
benchmark which produces as high as system-call rate as can be
expected under realistic conditions.

None of the usual embedded-system benchmarks produce significant
syscall loads on the microkernel, we therefore use a server-style
benchmark. Note that the nature of the benchmark is completely
irrelevant for this exercise, all that counts is the rate and
distribution of kernel entries. The relevant operations are IPC and
interrupt handling, as all other microkernel operations deal with
resource management that is relatively infrequent. 

The seL4 equivalent of a syscall in a monolithic system
is sending an IPC message to a server process and waiting for a reply
(i.e.\ two microkernel IPCs per monolithic OS syscall). Similarly, an
interrupt, which in a monolithic OS results in a single kernel entry,
produces two for the microkernel-based system, as the interrupt is
converted by the kernel into a notification to the driver (one kernel
entry), and the driver acknowledges to the kernel with another
syscall.

\subsubsection{Benchmark setup}\label{s:redis-expl}

In order to hammer our kernel, we use a simple client-server scenario,
consisting of the Redis key-value store \citep{redis:url}. We
consolidate the clients and servers on the same machine due to
insufficient network bandwidth to saturate the large number of cores.
Redis receives client requests from a virtual network processor on
Core~0. Each client and server has their own private copy of the lwIP
TCP/IP stack \citep{lwip:url} running as a usermode process.

\begin{figure}[htb]
  \centering
  \includegraphics[width=\columnwidth]{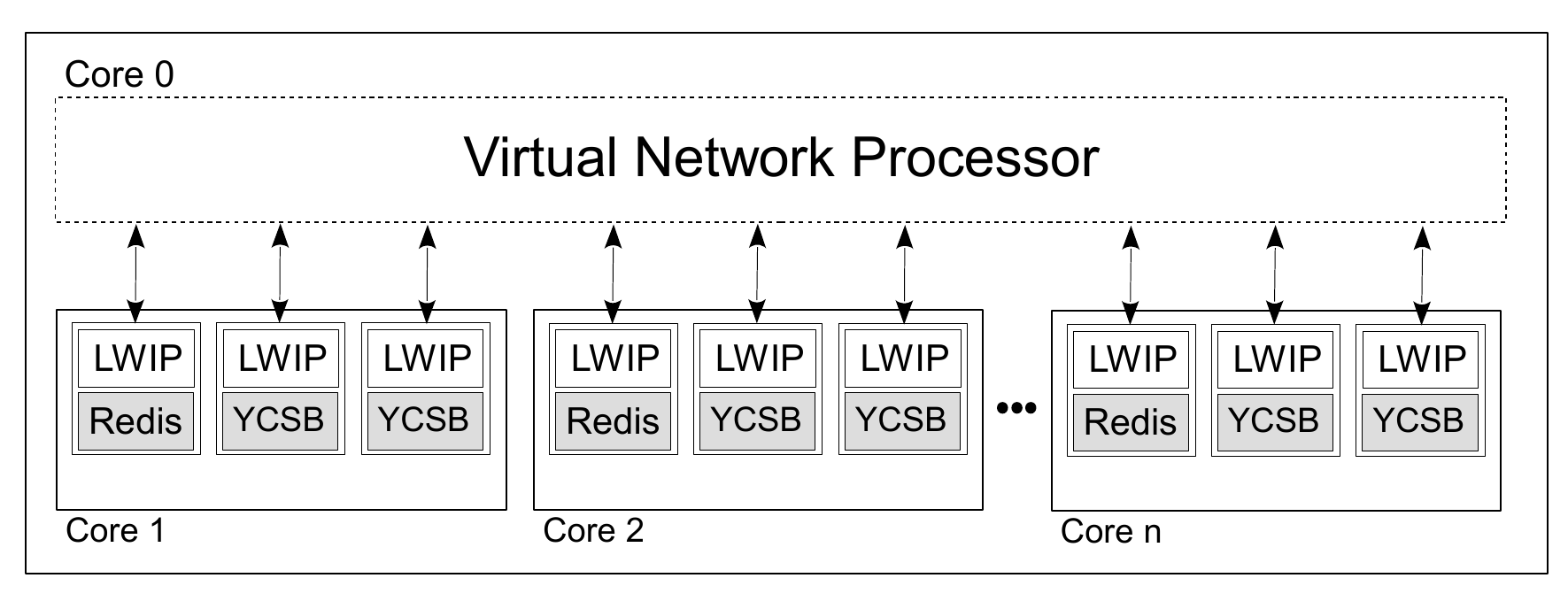}
  \caption{Redis-based benchmark architecture. }
  \label{fig:redis}
\end{figure}

\autoref{fig:redis} shows the system under test. With the exception of
Core~0, each core has a Redis server and two copies of the Redis
benchmarking client. 

We run Redis as volatile instances, i.e.\ disabling file-system
access, in order to maximise throughput and therefore the rate of
kernel entries.

We evaluate the performance using a modified version of Yahoo!~Cloud
Serving Benchmarks (YCSB) \citep{Cooper_STRS_10} as the benchmarking
client.  All client instances start simultaneously and are tuned to
perform a fixed number of operations that result in at least 2 minutes of
run time. The benchmarking client consists of the read-only workload
with zipfian distribution as presented in \citet{Cooper_STRS_10}. For
each kernel variant, we instrument the kernel to record idle time
within the idle loop for obtaining CPU utilisation for each run.

\subsubsection{Results}

\begin{figure}[htb]
  \centering
  \includegraphics[width=\columnwidth]{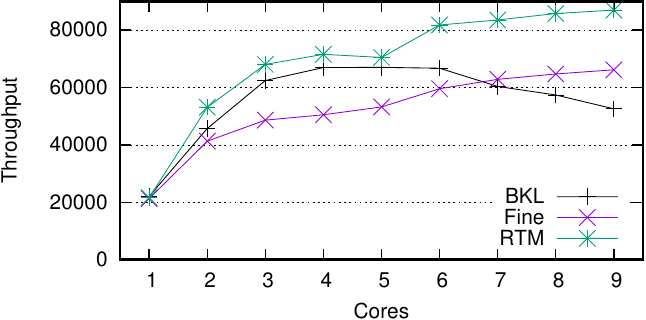}
  \caption{Redis-based benchmark architecture. }
  \label{fig:redis_tput}
\end{figure}

\autoref{fig:redis_tput} shows the results of the Redis benchmark on the x86 platform.\footnote{These are single runs, sorry, no standard deviations...} Not surprising, given the extreme workload, scalability is limited.

The transactions-based kernel performs consistently best. However, the BKL keeps up until 5 cores, after which throughput starts to drop. Fine-grained locking consistently performs at about 60-75\% of the transactions kernel.

We measure the average delay time on each core by dividing user time by the number of kernel entries. We find that the average time is 800 cycles on core 0 (the virtual network), and around 1,600 cycles on the client/server cores (figures taken from the five-core case). This confirms that the benchmark is fairly extreme in terms of system call load.

We can relate the observed delay times back to \autorefsub{fig:model}{fig:queueing}. With 800 cycles delay on one core and 1,600 on the others, we expect a behaviour similar to the 1,000-cycle delay curve of \autorefsub{fig:model}{fig:queueing}. And the similarity is indeed there: the 1,000-cycles curve peaks at 7 cores and then drops of slowly. The BKL curve of \autoref{fig:redis_tput} peaks slightly earlier but overall looks similar.

We re-iterate that this is an extreme benchmark, and \autorefsub{fig:model}{fig:queueing} tells us that a slightly less extreme version, with about four times the average delay, should scale to a full socket. 

\section{Related Work}
Writing parallel and scalable code is a topic almost as old as
computing itself. \citet{Cantrill_Bonwick_08} provide some historical
context and motivation for concurrent software, together with words of
wisdom to tackle the difficulties of writing high-performance and
correct concurrent software. We adhere to their advice by avoiding
parallelising complex software (i.e.\ splitting the BKL) as our data
shows it is unwarranted for closely coupled cores.

Recent complementary work evaluates the scalability of various synchronisation
primitives \citep{David_GT_13} on many-core processors. The authors
reinforce that scalability is a function of the hardware, with
scalability best when access is restricted to a single socket with
uniform memory access -- exactly our area of interest.

\citet{Boyd-Wickizer_KMZ_12} use queueing theory to predict the
collapse of ticket locks in Linux. We use a different model to predict
performance of a microkernel synchronised with a scalable lock, not
just the lock itself. We also use it to validate the scalability of
the implementation of the microkernel itself.

Hardware transactional memory is utilised in TxLinux
\citep{Rossbach_HPRAW_07} to implement \emph{cxspinlocks}, a
 combination of co-operative spinlocks and transactions
capable of supporting device I/O and nesting. A small microkernel
needs neither, as I/O is at user-level, and it can be designed to avoid
complex, nested, fine-grained locks. 

\citet{Hofmann_RW_09} apply HTM to coarse-grained locks in Linux 2.4
on a simulated HTM system. Our goals are similar, however our
experiments are on real hardware (Intel TSX), on a microkernel with a
single lock. Our event-based kernel avoids the need for a
\emph{cxmutex} to handle waking waiting threads on exiting a
transaction.

Patches for Linux to utilise Intel TSX have been made available
\citep{kleen:url}. To our knowledge, no performance data was
released. Eliding existing fine-grained locking does nothing to reduce
kernel complexity. We elide the whole microkernel, providing favourable
performance while retaining simplicity.

\section{Conclusions}\label{s:concl}

We have analysed scalability of a microkernel with fast system calls
across closely-coupled processor cores. We find that for such a
system, the overhead of locking is significant, ranging between a few
percent for a single lock on x86 to 23\% on an ARM processor. This
makes the best approach to concurrency control non-obvious, especially
when keeping in mind that it makes no sense to scale such a kernel to
a large multicore, where inter-core cache-line migration latencies
exceed the basic syscall cost.

We analysed three different multicore implementations of seL4, one
using a big kernel lock, one using fine-grained locking, and the third
using hardware-supported transactions. We evaluated the implementation
on a server-class x86 platform as well as an ARM multicore aimed for
the embedded market. We support the experiments with a
queueing-theoretical model.

There are three main take-aways from our evaluation. One is that the
inter-core cache-line migration cost matters a lot. This is
demonstrated by the ARM results, where the BKL scales perfectly to 4
cores even with the unrealistically high 500-cycle average
inter-syscall time. If architects can maintain similarly tight
coupling with higher core counts, our modelling shows that the BKL can be expected scale further. 

In contrast, the x86 platform shows that the perfect scaling regime
does not extend past two cores even with about double the
inter-syscall time, and plateaus at four cores (but keep in mind that
this is still an extremely, if not unrealistically high load). 

The second take-away is that lock overhead is significant on a well-designed microkernel with very short syscall latencies. This is particularly obvious on the ARM with its relaxed memory model and resulting high barrier costs, but the effect is also significant on the x86, where the kernel using fine-grained locking performs only at about 75\% of the BKL version until the latter reaches its performance knee.

The third takeaway is that hardware transactions are an exciting development. 
To our knowledge, we are the first to implement lock elision for a
BKL kernel using Intel's RTM. We show in microbenchmarks that a
theoretically embarrassingly parallel application scales perfectly
with little overhead and no serialisation. In our realistic (but extreme) macro
benchmark, which is less parallel, RTM upper-bounds the performance of both the BKL as well as fine-grained locking. In the case of the microkernel, where the whole system call can be packed into a single transaction, RTM gives get the best of both worlds.

We can summarise our experience that transactions are the way to go,
if they are available. Failing that, the big lock is actually a good
choice for a fast microkernel, as it is only outperformed by
fine-grained locking under extreme circumstances, at least on the kind
of closely-coupled system where a single, shared kernel instance makes
sense. Under those circumstances, the reduced (compared to
fine-grained locking) implementation complexity is a strong asset, as
it enables formal verification, which is presently unfeasible for
systems using fine-grained locking. The significantly better
performance under less extreme workloads is an added benefit.

\label{p:lastpage}
\balance
{\sloppy

  \ifAlpha
    \bibliographystyle{alpha}
  \else
    \bibliographystyle{plainnat}
  \fi
  \bibliography{references}
}
\end{document}